\documentclass[10pt,letterpaper,twocolumn]{article} 

\usepackage{ol2}
\usepackage[draft]{hyperref}
\usepackage{amsmath}

\begin{document}

\twocolumn[

\title{Tunable electromagnetic chirality induced by graphene inclusions in multi-layered metamaterials}

\author{Carlo Rizza $^{1,2}$, Elia Palange$^{2}$, Alessandro Ciattoni$^{3,*}$}

\address{
$^1$ Dipartimento di Scienza e Alta Tecnologia, Universit\`a  dell'Insubria, via Valleggio 11, I-22100 Como, Italy \\
$^2$ Dipartimento di Scienze fisiche e chimiche, Universit\`a  di L'Aquila, via Vetoio 1, I-67100 Coppito L'Aquila, Italy\\
$^3$ Consiglio Nazionale delle Ricerche, CNR-SPIN, via Vetoio 1, I-67100 Coppito L'Aquila, Italy\\
$^*$ Corresponding author: alessandro.ciattoni@aquila.infn.it}

\begin{abstract}
We theoretically investigate the electromagnetic response of a novel class of multi-layered metamaterials obtained by alternating graphene sheets and
dielectric layers, the whole structure not exhibiting a plane of reflection symmetry along the stacking direction. We show that the electromagnetic
response of the structure is characterized by a magneto-electric coupling described by an effective chiral parameter. Exploiting the intrinsic tunability
of the graphene-light coupling, we prove that one can tune both the dielectric and the chiral electromagnetic response by varying the graphene chemical
potential through external voltage gating.
\end{abstract}

\ocis{160.3918,230.4170,250.5403} ]

\noindent

Graphene, a one-atom thick layer of carbon atoms arranged in a honeycomb lattice, shows a wide range of unique properties \cite{Castro_Neto}. For example,
graphene exhibits high thermal and electric conductivity, high optical damage threshold and high third-order optical nonlinearities \cite{Bao}. Recently,
many graphene-based photonic and optoelectronic devices have been proposed and developed such as plasmonic waveguides \cite{Mikhailov_2,Jablan,Koppens},
frequency multipliers \cite{Mikhailov}, modulators \cite{Liu}, photodetectors \cite{Mueller} and polarizers \cite{Bao_2}. In the context of metamaterials,
A. Vakil \emph{et al.} \cite{Engheta} have theoretically proposed a setup where a graphene sheet is a one-atom-thick platform for achieving the desired
infrared metamaterials and transformation optical devices. On the other hand, several researchers have investigated multilayer structures composed of
stacked graphene sheets separated by thin dielectric layers \cite{Andryieuski,Iorsh,Othman,Othman_2}. A newsworthy advantage of such proposed
metamaterials is the overall tunability of the electromagnetic response which is entailed by the dependence of the graphene conducibility on the chemical
potential. For example, the graphene-based metamaterial response can be tailored from elliptic birefringent to hyperbolic by varying the graphene chemical
potential through an external gate voltage \cite{Iorsh}.

In this Letter, we propose a novel class of graphene-based metamaterials exhibiting a marked chiral electromagnetic response and we demonstrate that such
nonlocal effect can be tuned by varying the chemical potential of graphene sheets. More precisely, we consider propagation of transverse magnetic (TM)
waves through a multi-layer periodic structure not exhibiting a plane of reflection symmetry whose unit cell comprises $N$ layers of different dielectric
materials alternated with $N$ graphene sheets. Exploiting a suitable multiscale approach where the period to wavelength ratio is the small expansion
parameter, we obtain the constitutive equations describing the spatially nonlocal metamaterial response. Specifically we refine the standard effective
medium theory by deriving higher order contributions predicting, in particular, an overall medium chiral response for those layer thicknesses not fully
assuring homogenization. Generally, a reciprocal or chiral magneto-electric coupling is a consequence of the medium $3$D or $2$D chirality, namely the
underlying constituents (organic molecules, proteins, "meta-molecules", etc.) exhibit mirror asymmetry \cite{Landau}. Chirality can produce newsworthy
effects such as optical rotation and negative refraction \cite{Zouhdi}. On the other hand, the configuration we consider in this Letter has a $1$D
chirality which is in turn tunable due to the presence of the graphene sheets. It is worth stressing that usually considered bilayer metal-dielectric
structures \cite{Elser, Orlov} and graphene-based metamaterials (considered in Refs.\cite{Iorsh, Othman, Othman_2} where the metamaterial unit cell
consists of a graphene sheet placed on top of a dielectric material) show electromagnetic response strongly affected by second order spatial dispersion
which, however, does not yield electromagnetic chirality since the structure geometry admits plane of mirror symmetry.

Let us consider TM waves propagating in a graphene-based metamaterial whose underlying multilayered structure has a unit cell obtained by stacking along
the $z$-axis, $N$ graphene sheets separated by $N$-layers of different media of thicknesses $d_j$ ($j=1,2,3,..,N$) (see Fig.1 where the case $N=2$ is
reported). The electromagnetic field amplitudes ${\bf E}=E_x(x,z) \hat{\bf e}_x+E_z(x,z) \hat{\bf e}_z$, ${\bf H}=H_y(x,z) \hat{\bf e}_y$ associated with
monochromatic TM waves (the time dependence $\exp(-i \omega t)$ has been assumed where $\omega$ is the angular frequency) satisfy Maxwell's equations
\begin{eqnarray}
\label{Maxwell}
&&\partial_z E_x-\partial_x E_z=i \omega \mu_0 H_y, \nonumber \\
&&\partial_z H_y=i \omega \epsilon_0 \epsilon_x(z) E_x, \nonumber \\
&&\partial_x H_y=-i \omega \epsilon_0 \epsilon_z(z) E_z,
\end{eqnarray}
where $\epsilon_x$ and $\epsilon_z$ are the $x$-component and the $z$-component of the dielectric permittivity tensor, respectively and both are periodic
functions of period $d=\sum_{j=1}^{N} d_j$. Here, the $j$-th graphene sheet response is described by the surface conductivity $\sigma_j$ so that the
surface current $K_{x j}=\sigma_j E_x$ yields a delta-like contribution to $\epsilon_x$ which is $i \sigma_j /(\omega \epsilon_0) \delta (z-z_j)$, $z_j$
being the sheet position \cite{Othman} (see below). Note that the surface conductivities $\sigma_j$ can assume different values in order to encompass the
relevant situation where the graphene can be locally tuned or substituted with more general bi-dimensional hetero-structures.

In order to obtain an effective medium description of the electromagnetic propagation in the regime where the ratio between the period $d$ and
the wavelength $\lambda$ is small, we exploit a standard and rather general multiscale technique \cite{Rizza,Rizza_2} holding for very general
$\epsilon_x(z)$ and $\epsilon_z(z)$ periodic profiles (which we will later specialize to considered graphene-based multi-layer). Accordingly we
introduce the parameter $\eta=d/\lambda \ll 1$ and the fast coordinate $Z=z/\eta$ and, aimed at isolating the slowly and rapidly varying
contributions, we consider the Fourier series of $\epsilon_x$ and $\epsilon_z^{-1}$, namely $\epsilon_x=\left \langle \epsilon_x \right
\rangle+\delta \epsilon_x$, $\epsilon_z^{-1}=\left \langle \epsilon_z^{-1} \right \rangle+\delta \epsilon_z^{(-1)}$ where $\left \langle f
\right \rangle$ is the mean value of the function $f$ and
\begin{eqnarray} \label{deps}
\delta \epsilon_x       &=&  \sum_{n\neq0} a_n \exp\left( i n \frac{ 2 \pi z}{d} \right) = \sum_{n\neq0} a_n \exp \left (i n k_0 Z\right), \nonumber \\
\delta \epsilon_z^{(-1)}&=&  \sum_{n\neq0} b_n \exp\left( i n \frac{ 2 \pi z}{d} \right) = \sum_{n\neq0} b_n \exp \left (i n k_0 Z\right). \nonumber \\
\end{eqnarray}
\begin{figure}[h!]
\includegraphics[width=1\linewidth]{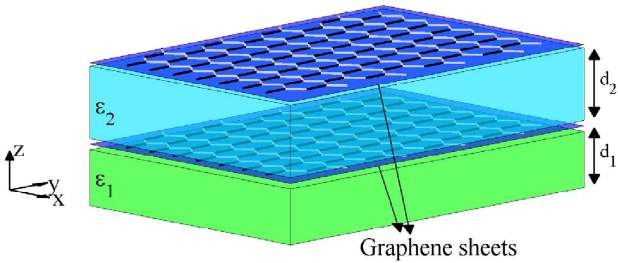}
\caption{(Color online) Sketch of graphene-based metamaterial unit cell. $\epsilon_j$ and $d_j$ ($j=1,2$) are the relative dielectric permittivities and
the thicknesses of the dielectric layers, respectively.}
\end{figure}
where $k_0 = 2\pi / \lambda$. The basic Ansatz of our approach is given by
\begin{eqnarray}
\label{fields}
E_x(x,z,Z) &=& \bar{E}_x(x,z) + \eta \delta E_x (x,z,Z), \nonumber \\
E_z(x,z,Z) &=& \bar{E}_z(x,z) + \delta E_z (x,z,Z), \nonumber \\
H_y(x,z,Z) &=& \bar{H}_y(x,z) + \eta \delta H_y (x,z,Z),
\end{eqnarray}
where $\bar A(x,z)$ and $\delta A (x,z,Z)$ are the slowly (averaged) and rapidly varying part of each electromagnetic field component $A$,
respectively ($A=E_x,E_z,H_y$). The considered Ansatz, where each field component is a Taylor expansion up to first order in $\eta$, has been
suitably chosen to self-consistently assure that finite and not trivial results are obtained in the asymptotic $\eta \rightarrow 0$ limit.
Substituting the Fourier series of $\epsilon_x$ and $\epsilon_z^{-1}$ and the Ansatz of Eqs.(\ref{fields}) into Maxwell equations
(\ref{Maxwell}), after separating the slowly and rapidly varying contributions, we obtain the coupled equations
\begin{eqnarray}
\label{average}
&& \partial_z \bar{E}_x-\partial_x \bar{E}_z=i \omega \mu_0 \bar{H}_y, \nonumber \\
&& \partial_z \bar{H}_y = i \omega \epsilon_0
\left( \left \langle \epsilon_x \right \rangle \bar{E}_x+\eta \left \langle \delta \epsilon_x \delta E_x \right \rangle \right), \nonumber \\
&& \bar{E}_z=\frac{i}{\omega \epsilon_0 }\left(\left \langle \epsilon_z^{-1} \right \rangle \partial_x \bar{H}_y +  \eta \left \langle \delta
\epsilon_z^{(-1)} \partial_x \delta {H}_y \right \rangle \right)
\end{eqnarray}
and
\begin{eqnarray}
\label{varying}
& \partial_Z\delta E_x-\partial_x\delta E_z=0, &\nonumber \\
& \partial_Z\delta H_y=i \omega \epsilon_0 \delta \epsilon_x \bar{E}_x, \quad \delta E_z=\frac{i}{\omega \epsilon_0 } \delta \epsilon_z^{(-1)}
\partial_x \bar{H}_y.&
\end{eqnarray}
It is important stressing that no terms have been neglected when deriving Eqs.(\ref{average}) whereas only the leading contributions (the lowest powers of
$\eta$) has been retained to obtain Eqs.(\ref{varying}). After integration on $Z$ and using Eqs.(\ref{deps}), Eqs.(\ref{varying}) yield the rapidly
varying parts of the field amplitudes as functions of the slowly ones, i.e.
\begin{eqnarray}
\label{sol-var}
&& \delta E_x =\frac{1}{k_0  \omega \epsilon_0} \sum_{n\neq0}  \frac{b_n}{n} e^{i k_0 n Z}    \partial_x^2 \bar{H}_y,     \nonumber \\
&& \delta E_z=-\frac{1}{i \omega \epsilon_0 } \sum_{n\neq0} b_n e^{i k_0 n Z}  \partial_x \bar{H}_y,   \nonumber \\
&& \delta H_y= \frac{ \omega \epsilon_0 }{k_0} \sum_{n\neq0} \frac{a_n}{n} e^{i k_0 n Z} \bar{E}_x.
\end{eqnarray}
Finally, substituting Eqs.(\ref{sol-var}) into Eqs.(\ref{average}), we get
\begin{eqnarray}
\label{Maxwell_eff}
&&\partial_z \bar{E}_x-\partial_x \bar{E}_z= i \omega \mu_0 \bar{H}_y, \nonumber \\
&&\partial_z \bar{H}_y= i \omega \epsilon_0 \left( \epsilon_x^{(eff)}  \bar {E}_x
     - i  \frac{\chi^{(eff)} Z_0 }{\epsilon_z^{(eff)} k_0^2} \partial_x^2 \bar{H}_y \right), \nonumber \\
&& \partial_x \bar{H}_y=-i \omega \epsilon_0  \left(\epsilon_z^{(eff)} \bar{E}_z + \frac{\chi^{(eff)}}{k_0}
\partial_x \bar{E}_x \right),
\end{eqnarray}
where $Z_0=\sqrt{\mu_0/\epsilon_0}$ is the vacuum impedance, and  $\epsilon_x^{(eff)}=\left \langle \epsilon_{x} \right \rangle$,
$\epsilon_z^{(eff)}=\left \langle \epsilon_{z}^{-1} \right \rangle ^{-1}$ and
\begin{equation}
\label{gamma} \chi^{(eff)}=i \eta \epsilon_z^{(eff)} \sum_{n \neq 0} \frac{a_{-n} b_{n}}{n}.             
\end{equation}
It is evident that in the limit $\eta \rightarrow 0$ the parameter $\chi^{(eff)}$ vanishes and the multiscale approach considered in this Letter
reproduces the results of the well known standard effective medium theory (EMT) \cite{Elser}. Furthermore, it is worth noting that in the case where the
structure admits mirror symmetry with respect a specific plane $z=z_0$, i.e. the relations $\epsilon_x(z) = \epsilon_x(-z+z_0)$ and $\epsilon_z(z) =
\epsilon_z(-z+z_0)$ hold, it is straightforward proving that the dielectric Fourier coefficients are such that $a_{-n}=\exp(i 2\pi n z_0/d) a_n$ and
$b_{-n}=\exp(i 2\pi n z_0/d) b_n$ so that $a_{-n}b_{n}=a_{n}b_{-n}$ and the series of Eq.(\ref{gamma}) provides a vanishing $\chi^{(eff)}$. Therefore, the
slowly varying and leading electromagnetic field can experience the effect of the novel terms proportional to $\chi^{(eff)}$ in the effective Maxwell
equations of Eq.(\ref{Maxwell_eff}) only if the multi-layer does not exhibit an inversion center i.e. if it is chiral. Comparing the second and the third
of Eqs.(\ref{Maxwell_eff}) with the standard equations $\partial_z \bar{H}_y= i \omega \bar{D}_x$, $\partial_x \bar{H}_y=-i \omega \epsilon_0 \bar{D}_z$
and using the third of Eqs.(\ref{Maxwell_eff}) to substitute for the magnetic field derivative we obtain
\begin{eqnarray} \label{consti}
\bar{D}_x &=& \epsilon_0 \left( \epsilon_x^{(eff)} \bar{E}_x - \frac{\chi^{(eff)}}{k_0} \partial_x \bar{E}_z - \frac{\chi^{(eff)2}}{\epsilon_z^{(eff)}
k_0^2} \partial_x^2 \bar{E}_x \right) \nonumber \\
\bar{D}_z &=& \epsilon_0 \left( \epsilon_z^{(eff)} \bar{E}_z + \frac{\chi^{(eff)}}{k_0} \partial_x \bar{E}_x \right).
\end{eqnarray}
which are the structure effective constitutive relations. Note that Eqs.(\ref{consti}) contain term proportional to the first and second $x$-spatial
derivative of the field components, term usually arising when dealing with weakly spatially nonlocal medium. Exploiting the fact that the effective
Maxwell's equations are invariant with respect to transformation $\bar{D}'_x=\bar{D}_x-\partial_z Q$, $\bar{D}'_z=\bar{D}_z+\partial_x Q$ and
$\bar{H}'_y=\bar{H}_y-i\omega Q$ (where $Q(x,z)$ is an arbitrary function), after setting $Q = - \epsilon_0 \chi^{(eff)}/k_0 \bar{E}_x$, we obtain the
equivalent effective constitutive relations
\begin{eqnarray}
\label{const} \bar{D}'_x&=& \epsilon_0 \left[ (\epsilon_x^{(eff)}+ \chi^{(eff)2}) \bar{E}_x - \frac{\chi^{(eff)2}}{\epsilon_z^{(eff)} k_0^2}
              \partial_x^2 \bar{E}_x \right] \nonumber \\
                        &+& i \frac{\chi^{(eff)}}{c}\bar{H}'_y, \nonumber \\
\bar{D}'_z&=& \epsilon_0 \epsilon_z^{(eff)} \bar{E}_z, \nonumber \\
\bar{B}_y &=& \mu_0 \bar{H}'_y-i \frac{\chi^{(eff)}}{c} E_x.
\end{eqnarray}
Therefore, in the limit $\eta \ll 1$ of quasi-homogenization regime, the effect of the multi-layer mirror asymmetry can be reinterpreted to yield an
effective chiral magneto-electric coupling (through the terms proportional to $\chi^{(eff)}$ in Eqs.(\ref{const})), a correction $\chi^{(eff)2}$ to the
$x$-component of the dielectric permittivity and a second order dispersion effect (see terms proportional to the second-order derivatives of electric
field in the first of Eqs.(\ref{const})).

In the contexts of chiral multilayers, graphene can play a twofold significant role since its sheets can turn a standard achiral structure into a chiral
one whose response, described by the above theory, can be further tuned by varying the graphene chemical potential. In order to discuss this point, we
consider a bilayer structure whose unit cell comprises two dielectric layers separated by a graphene sheet. The dielectric permittivities of such
structure can be written, within the unit cell $0 \leq z < d$ as $\epsilon_x=\Xi(z)+\frac{i \sigma_1}{\omega \epsilon_0 } \delta(z-d_1)$ and
$\epsilon_z=\Xi(z)$ where $\Xi(z)=\epsilon_1 \Pi \left( \frac{z-d_1/2}{d_1} \right) +\epsilon_2\Pi \left( \frac{z-d_2/2-d_1}{d_2} \right)$, $\Pi(z)$ is
the rectangular function ($\Pi(z)=0$ if $|z|>1/2$, $\Pi(z)=1/2$ if $|z|=1/2$, $\Pi(z)=1$ if $|z|<1/2$), $\delta(z)$ is the Dirac delta function and
$\sigma_1$ is the surface conductivity of the graphene layer. In this model, the graphene sheet is infinitesimally thin and the current it supports is
along the $x$-direction thus solely affecting the bilayer $x$-component of the permittivity tensor. Note that the considered structure is chiral since the
permittivity component $\epsilon_x(z)$ does not show along the stacking direction a plane of mirror symmetry and it is worth stressing that this due to
the graphene sheets. The structure effective parameters are easily evaluated and are
\begin{eqnarray}
\label{x_z_chi} &\epsilon_x^{(eff)} = \frac{1}{d} \left ( d_1 \epsilon_1+d_2 \epsilon_2 + i \frac{\sigma_1}{\omega \epsilon_0} \right),
\quad \epsilon_z^{(eff)} = d \left (\frac{d_1}{\epsilon_1} +\frac{d_2}{\epsilon_2} \right)^{-1}, & \nonumber \\
& \chi^{(eff)} =  i \frac{d_1 d_2}{2 c \epsilon_0 d^2} \epsilon_z^{(eff)} \sigma_1 \left( \frac{1}{\epsilon_2} - \frac{1}{\epsilon_1} \right),&
\end{eqnarray}
where the expression of $\chi^{(eff)}$ is obtained after the straightforward summation of the series in Eq.(\ref{gamma}). Evidently $\chi^{(eff)}$
vanishes if there is no graphene ($\sigma_1=0$, discussed in Refs.\cite{Elser, Orlov}) or if the dielectrics are identical ($\epsilon_1=\epsilon_2$
discussed in Refs.\cite{Iorsh, Othman, Othman_2}) since in both situations the structure is achiral. In the following numerical examples, we choose the
wavelength $\lambda=10.71$ $\mu$m and the layer dielectric permittivities $\epsilon_1=-1.87 + 0.16i$, $\epsilon_2=2.25$ associated to silicon carbide
(SiC) \cite{Palik} and PMMA \cite{Graf}, respectively. In addition we adopt the semiclassical expression for the graphene conductivity $\sigma_1$ holding
if $|\mu_c| \gg K_b T$ ($\mu_c$ is the chemical potential, $K_b$ is the Boltzmann's constant and $T$ is the temperature) and obtained by taking into
account the inter- and intra-band contributions (see Eqs.(4) and (5) in Ref.\cite{Hanson}). Note that the graphene surface conductivity depends on the
frequency $\omega$, the chemical potential $\mu_c$, the temperature $T$ and the phenomenological scattering rate $\Gamma$. Here we assume $T=300$ K and
$\Gamma=0.43$ meV. In addition, setting $d_1=d_2$, the effective permittivity $z$-component is $\epsilon_z^{(eff)}=-18.43 + i 9.60$ and it is not affected
by the chemical potential, whereas the effective permittivity $x$-component $\epsilon_x^{(eff)}$ and the chiral parameter $\chi^{(eff)}$ can be tuned by
varying the graphene chemical potential through external voltage gating.

In Fig. 2, we report the real (solid line) and imaginary (dashed line) parts of $\epsilon_x^{(eff)}$ and $\chi^{(eff)}$, respectively, as functions of
$\mu_c$ for $\eta=1/15$. The tunability of the overall electromagnetic response is evident and it also remarkable that in this case a transition from
hyperbolic behavior to anisotropic negative dielectric one occurs: the real part of the $x$-component of the dielectric permittivity $\epsilon_x^{(eff)}$
is positive in the region $0.1$ eV $<\mu_c<$ $0.2$ eV (shadow area in Fig.2(a)) and it is negative in the region $\mu_c>0.2$ eV, whereas
$Re(\epsilon_z^{(eff)})$ is negative everywhere.
\begin{figure}[h!]
\centering
\includegraphics[width=0.9\linewidth]{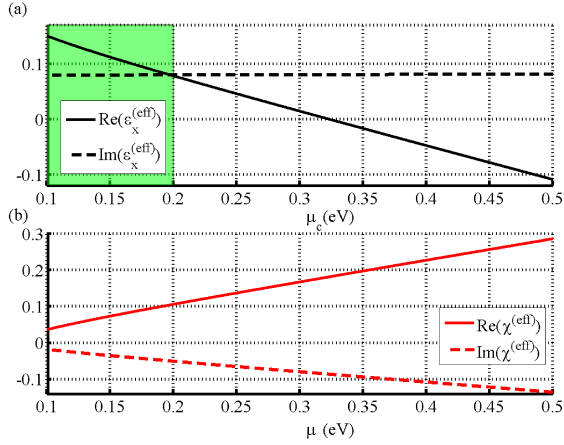}
\caption{(Color online) Effective parameters (a) $\epsilon_x^{(eff)}$ and (b) $\chi^{(eff)}$  as functions of the graphene chemical potential $\mu_c$.}
\end{figure}

In order to check and discuss the predictions of our multiscale approach, we here consider the scattering process of TM waves by a graphene-based
metamaterial slab lying in the region $0< z <L$ which, using the standard transfer matrix-method, admits full analytical description. Accordingly we
evaluate the exact optical transfer function defined as $OTF={H}^{(t)}_y/ {H}^{(i)}_y$ where ${H}^{(i)}_y$, ${H}^{(t)}_y$ are the amplitude of the
incident and transmitted magnetic field evaluated at $z=0$ and $z=L$, respectively and where the $x$ dependence $\exp (i k_x x)$ has been assumed ($k_x$
is the transverse component of the wave vector). On the other hand, the system of Eqs.(\ref{Maxwell_eff}) together with Eqs.(\ref{sol-var}) can be solved
to obtain the OTF in the quasi-homogenized regime. In Fig.3 we compare the exact OTF (solid lines) with those predicted by our multiscale approach (dash
lines) and by the standard EMT (dash-circle lines) for $L=4.01$ $\mu$m, $\eta=1/8$, $\mu_c=0.3$ eV. In this example, the $x$-component of dielectric
permittivity and the chiral coefficient are $\epsilon_x^{(eff)}= 0.095 + i 0.079$, $\chi^{(eff)}= 0.17 - 0.079i$, respectively. We note that our non-local
multiscale approach is in good agreement with the exact OTF (both predict a resonance at $k_x=1.55 k_0$ as shown in Fig.3(a)), whereas the standard EMT is
not adequate to describe the realistic situation.
\begin{figure}[h!]
\includegraphics[width=1\linewidth]{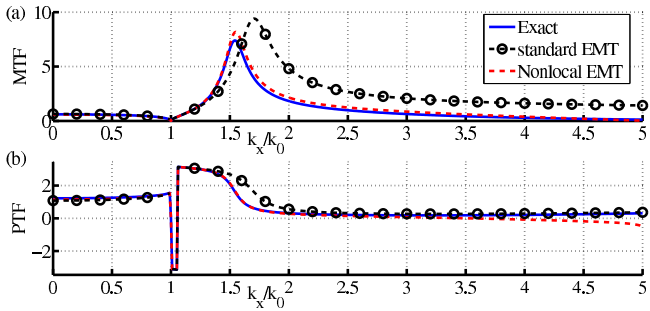}
\caption{(Color online) Comparison among the OTFs as evaluated from the exact matrix-method (solid lines), from the standard EMT (dash-circle lines) and
from the non-local multiscale EMT (dash lines). The functions MTF and PTF are related to the OTF by the relation $OTF(k_x)=MTF(k_x) \exp\left[i
PTF(k_x)\right]$.}
\end{figure}

In conclusion we have shown that a multilayer structure not exhibiting mirror symmetry along the stacking direction (chiral structure), in the
quasi-homogenized regime, provide first-order electromagnetic nonlocal response. In particular, we have argued that graphene sheets suitably
inserted within a achiral structure can turn it into a chiral one whose nonlocal electromagnetic response is tunable through the graphene
chemical potential.

This research has been funded by the Italian Ministry of Research (MIUR) through the "Futuro in Ricerca" FIRB-grant PHOCOS - RBFR08E7VA and by Progetto
DOTE Lombardia.

\newpage
\pagebreak

\end{document}